\documentclass{article}
\def\beq{\begin{equation}}
\def\eeq{\end{equation}}
\def\bea{\begin{eqnarray}}
\def\eea{\end{eqnarray}}
\def\bq{\begin{quote}}
\def\eq{\end{quote}}

\def\rar{\rightarrow}

\def\la{\langle}
\def\ra{\rangle}

\def\b1{\beta_1}

\begin{document}
\topmargin -1.5cm
\oddsidemargin +0.2cm
\evensidemargin -1.0cm
\pagestyle{empty}
\begin{flushright}
PM/97-32\\ 
\end{flushright}
\vspace*{5mm}
\begin{center}
\section*{Direct extraction of the chiral quark condensate\\ 
and bounds on the light quark masses}
\vspace*{0.5cm}
{\bf H.G. Dosch}\\
\vspace{0.3cm}
Institut f\"ur Theoretische Physik der Universit\"at\\
Philosophenweg 16, D69120 Heidelberg, FRG.\\
E-mail: h.g.dosch@thphys.uni-heidelberg.de\\
\vspace{0.3cm}
and\\
\vspace{0.3cm}
{\bf S. Narison}
\\
\vspace{0.3cm}
Laboratoire de Physique Math\'ematique et Th\'{e}orique\\
UM2, Place Eug\`ene Bataillon\\
34095 - Montpellier Cedex 05, France\\
E-mail:
narison@lpm.univ-montp2.fr\\
\vspace*{1.5cm}
{\bf Abstract} \\ \end{center}
\vspace*{2mm}
\noindent
We select sum rules from which one can extract directly reliable
limits on the size of the
chiral symmetry breaking light quark condensate  $\la\bar\psi\psi\ra$.
Combined results from the nucleon and $B^*$-$B$ mass-splitting sum rules
give a result compatible with the standard value: 
$\la\bar\psi\psi\ra$(1 GeV)$\simeq
(-229~ {\rm MeV})^3$, through the determinations of the
quark-gluon mixed condensate. The vector form factor of the $D\rar
K^*l\nu$ semi-leptonic decay leads to the range 
$0.6\leq  \la\bar\psi\psi\ra/ (-229 {\rm ~MeV})^3 \leq 1.5$. The upper
limit combined with the Gell-Mann-Oakes-Renner (GMOR) relation implies the
interesting lower  bound on the sum of light quark masses:$(m_u+m_d)$(1
GeV)$\geq$ 9.4 MeV, which combined with the ratio of light quark masses from
chiral perturbation theory leads to $m_s$(1 GeV)$\geq (121\pm 12)$ MeV. The
lower limit combined with the positivity of the $m_q^2$ contribution to the GMOR
relation leads to the upper bound : $(m_u+m_d)$(1 GeV)$\leq$ 
15.7 MeV, which is independent on the nature of chiral symmetry breaking.

\noindent
 
\vspace*{3cm}
 
\begin{flushleft}
PM/97-32 \\
August 1997
\end{flushleft}
\vfill\eject

\pagestyle{plain}
\setcounter{page}{1}
\section{Introduction}
The chiral condensate of light quarks $ \la\bar\psi\psi\ra$ is one of the
fundamental 
parameters of non-perturbative QCD and chiral symmetry. Therefore it is of
prime interest to determine it, as directly as
possible, from experimental data . It is related to the pion decay constant
$f_\pi=93.3$ MeV, the pion mass
$m_\pi$ and the sum of light quark mass $(m_u + m_d)$ by the
Gell-Mann Oakes Renner (GMOR) relation \cite{GMOR68}: 
\begin{equation} \label{GMOR}
m_\pi^2 f_\pi^2 = -(m_u+m_d) \la\bar\psi\psi\ra + O(m_q^2)
\end{equation}
In the standard treatment of chiral symmetry breaking 
(chiral perturbation theory \cite{LEUT}), the light quark
masses are very small ($m_q \leq 10$ GeV) and therefore the $O(m_q^2)$
terms are negligible. In other approaches 
(generalized chiral perturbation theory \cite{STERN}) the quark
masses are not so small and therefore the $O(m_q^2)$ term might become
important or even dominant.
A precise determination of the condensate  is therefore of great
theoretical interest for clarifying the nature of the mechanism of chiral
symmetry breaking. In the standard approach, the  GMOR-relation
(Eq. \ref{GMOR}) and the value of the quark condensate also allow a
determination of the absolute values of the light quark masses $u$ and $d$
independently of other sum rules approaches, such as the pseudoscalar
\cite{PSEU,PSEUDO,EDUAR} for $(m_u+m_d)$ and the scalar sum rules for $(m_d-m_u)$
\cite{SCALAR},  and the present lattice extrapolations \cite{LATT}. 
\section{The chiral condensate from QCD sum rules.}
The chiral condensate plays an important role in the QCD sum rule 
\cite{SVZ79} analysis of many channels and the standard value
\cite{MS,SNB,PSEUDO} \footnote{The errors quoted here and in the following
come from the variations of the different input parameters and of the
sum rule variables at given order of perturbation theory.}:
\begin{equation}\label{standard}
 \la\bar\psi\psi\ra(1~\rm {GeV})= - [(229 \pm 9){\rm MeV}]^3
\end{equation}
has led to interesting results \cite{SNB}, many of which have
been checked experimentally.
However, if one wants to extract the value of $ \la\bar\psi\psi\ra$
directly from phenomenological data, correlations with other (non)
perturbative parameters limit the accuracy severely. In the light meson sum
rules, the chiral quark condensate effects are relatively negligible
compared with the ones of the gluon condensate $\la \alpha_s G^2 \ra$, as
it is often multiplied by the small light quark mass. 
In the nucleon sum rules \cite{CDKS81,Iof81,DJN89,
SNB}, which seem, at first sight, a very good
place for determining $ \la\bar\psi\psi\ra$,  we have two
form factors for which spectral sum rules can be constructed, namely
the form factor $F_1$ which is proportional to the Dirac matrix
$\gamma\,p$ and $F_2$ which is proportional to the unit matrix. In
$F_1$ the four quark condensates play an important role, but these are
not chiral symmetry breaking and are related to the condensate $
\la\bar\psi\psi\ra$ only 
by the factorization
hypothesis \cite{SVZ79} which is known to be violated by a factor 2-3 
\cite{CDKS81,TARRACH,SNB}.
The form factor $F_2$ is dominated by the condensate $
\la\bar\psi\psi\ra$ and the mixed 
condensate $\la\bar{\psi} \sigma G \psi\ra$, such that the baryon mass is
essentially determined by the ratio $M_0^2$ of the two condensates:
\begin{equation}\label{mixed}
M_0^2 = { \la\bar{\psi} \sigma G \psi\ra}/{ \la\bar\psi\psi\ra}
\end{equation}
Therefore from nucleon sum rules one  gets a rather reliable
determination of $M_0^2$ \cite{Iof81,DJN89}:
\begin{equation} \label{m0}
M_0^2 = (.8\pm .1)~\rm{GeV}^2.
\end{equation}
A sum rule based on the ratio $F_2/F_1$ would in principle be
ideally suited for a determination of $ \la\bar\psi\psi\ra$ but
this sum rule is
completely unstable \cite{DJN89} due to fact that odd parity
baryonic
excitations contribute with different signs to the spectral
functions
of $F_1$ and $F_2$.
In the correlators of heavy mesons ($B,B^*$ and $D,D^*$) the chiral
condensate gives a significant direct contribution in contrast to
the light meson sum rules \cite{SNB}, since, here, it is multiplied by the
heavy quark mass. However, the dominant contribution to the meson mass
comes from the heavy quark mass and therefore a change of a factor
two
in the value of $ \la\bar\psi\psi\ra$ leads only to a negligible
shift of the mass. However, from the $B$-$B^*$ splitting one gets
a precise determination of the mixed condensate 
$\la\bar{\psi} \sigma G \psi\ra $
with the value \cite{BB*}
\begin{equation}\label{mixed2}
\la\bar{\psi} \sigma G \psi\ra =  -(9\pm 1)\times 10^{-3}~\rm{GeV^5}~,
\end{equation}
which combined with the value of $M_0^2$ given in Eq. (\ref{m0}) gives
our first result for the value of $ \la\bar\psi\psi\ra$
at the nucleon scale: 
\begin{equation}\label{result1}
 \la\bar\psi\psi\ra(M_N)= -[(225 \pm 9)~\rm{MeV}]^3
\end{equation}
in good agreement with the standard value in Eq. (2).\\
Other good channels for the determination of $
\la\bar\psi\psi\ra$ are the
sum rules for the semileptonic decays of heavy pseudoscalars to
light vector mesons \cite{BBDN91,BBD91,BRHO}, where the chiral
condensate plays
a dominant role. For the decay $D\, \to \, K^*\ell\nu$ there exist
now rather precise data for the semileptonic form factors. In a
previous analysis \cite{BBDN91,BBD91} the range of the exponential sum rules
parameters
$\tau_1\, , \tau_2$ of the three-point function was fixed by the
corresponding two point 
functions and the choice of standard parameters allowed successfull
predictions for the form factors. In this analysis where we want to
extract from experiment the fundamental quantity $\la \bar \psi\psi\ra$,
we do not make a preselection and choose the values of $\tau_1$ and
$\tau_2$ by requiring stability of
the three-point function in both (uncorrelated) parameters. 
Making use of the standard definition of the semileptonic decay
form
factors (see e.g. PDG 96 \cite{PDG}) we find that the above-mentioned
stability criterion is only fullfilled for the vector form factor
$V$. For that
quantity at zero momentum transfer, we obtain the sum rule:
\begin{eqnarray}\label{srv}
V(0) &=& \frac{m_c(m_D+m_{K^*})}{4 f_D f_{K^*} m_D^2 m_{K^*}}
exp[(m_D^2-m_c^2)\tau_1 + m^2_{K^*} \tau_2]\\
 &\times&\la\bar\psi\psi\ra\Bigg{\{} - 1 + M_0^2(-\frac{\tau_1}{3} +
\frac{m_c^2}{4}
\tau_1^2 +\frac{2 m_c^2- m_c\,m_s}{6}\tau_1\tau_2)\nonumber \\
&&-\frac{16 \pi}{9} \alpha_s \rho  \la\bar\psi\psi\ra (\frac{2 m_c}{9} 
\tau_1 \tau_2
-
\frac{m_c^3}{36} \tau_1^3 \nonumber \\
&&- \frac{2 m_c^3-m_c^2 m_s}{36}\tau_1^2 \tau_2
+\frac{-m_c}{9}\tau_1^2 + \frac{2 m_s}{9}\tau_2^2
+\frac{2}{9}m_s\tau_1\tau_2+
\frac{4}{9}\frac{\tau_2}{m_c})\nonumber \\
&& + \frac{e^{m_c^2 \tau_1}}{
\la\bar\psi\psi\ra}\int_0^{s_{20}}ds_2
\int_{s_2+m_c^2}^{s_{10}}ds_1\, \rho_v(s_1,s_2) e^{-s_1 \tau_1 - s_2
\tau_2}\Bigg{\}}
\nonumber \\
{\rm with}&& \rho_v(s_1,s_2) = \frac{3}{4 \pi^2 \, (s_1-s_2)^3} \times
\\ &&\qquad \Big{\{}m_s((s_1+s_2)(s_1-m_c^2) - 2 s_1 s_2) + m_c((s_1 +
s_2) s_2 - 2 s_2
(s_1-m_c^2))\Big{\}} \nonumber
\end{eqnarray}
The factor $\rho$ expresses the uncertainty in the factorization of
the four quark condensate. In our numerical analysis,
we start from the following set of standard parameters given in Eqs. (2)
and (4) and: $f_{K^*} = 0.15~ {\rm GeV} (f_\pi=93.3~\rm{MeV}),~
 m_c({\rm pole}) = (1.42 \pm 0.02)~ {\rm GeV}$,
where the value of the charm quark mass comes from \cite{SNM}. 
We enlarge the value of the charm pole mass until
half of $M_{J/\psi}$ (so-called constituent mass)
i.e $m_c=1.42\pm 0.13$ GeV, in order to be conservative.
We use the recent value of the four-quark condensate correlated to
the value of the gluon condensate obtained in \cite{SNGLU},
which is consistent with the value $\rho=2\sim 3$ for the condensate
value in Eq. (2). The value of $f_D\simeq (1.35\pm 0.07)f_\pi$ is consistently
determined  by a two-point function sum rule including radiative corrections, where
the sum rule expression can, e.g., be found in \cite{SNB}.
The following parameters enter only marginally:
$m_s(1~{\rm GeV}) = (0.15\sim 0.19)~{\rm GeV},~ s_{10} = (5\sim 7)~
{\rm GeV}^2,~ s_{20} = (1.5\sim 2)~{\rm GeV}^2$.
With the previous inputs, we obtain a stationary point in the two variables
$\tau_1$ and $\tau_2$ at the values 0.8 and 0.9 GeV$^{-2}$
respectively and $V(0) = 1.12$ which agrees well with the
experimental value 
$V(0) = (1.1 \pm .2$) \cite{PDG}.
Normalized to $\la\bar\psi\psi\ra$, the relative
contributions of the other terms are respectively:
-0.32 for the
mixed condensate, +0.01 for the four quark condensate and
0.25 for the perturbative one, thus showing that at these
values of the
$\tau$-variables the local OPE converges quite well, and appears to be
in a good control.
Now, we use the experimental value $V(0)= (1.1\pm 0.2)$.
In this way, we obtain the following range of values for the chiral
condensate evaluated at the $\tau$-sum rule scale of 0.8-0.9 GeV$^{-2}$:
\begin{equation}\label{main}
0.6 \leq {\la\bar\psi\psi\ra}/{(-0.229\,{\rm GeV})^3} \leq 1.5,
\end{equation}
where the lower (resp. higher) bound corresponds to the small
(resp. big) value of the charm quark mass.
 Releasing the absolute value of
the mixed condensate and keeping the ratio $M_0^2$ fixed would weaken the
upper limit in Eq. (\ref{main}) to 2.2 since the mixed condensate has the
opposite
sign of the quark condensate.
The result in Eq. (\ref{main}) of the analysis of the semileptonic decay is
thus
in perfect 
agreement with the result in Eq. (\ref{result1}) of the combined nucleon and
$B$-$B^*$ splitting analysis. The previous lower bound
excludes smaller values  of ${\la\bar\psi\psi\ra}$ which may appear within
the framework of generalized chiral perturbation theory \cite{STERN} and
questions the reliability of the recent estimate based on variational
approach within perturbation theory \cite{KNEUR}. 
\section{Bounds and values of the light quark masses}
The previous upper limit is specially interesting since it gives, through
the GMOR relation in Eq. (\ref{GMOR}), also a lower limit on the sum of light
quark masses:
\begin{equation}
{(m_u+m_d)(1~\rm GeV)} \geq 9.4~ {\rm MeV}. 
\end{equation}
The lower bound in Eq. (9) can be exploited by using the positivity of the 
$m^2_q$ term \cite{STERN}, which could be dominant in the generalized chiral
perturbation theory approach. In this way, one can also derive the upper limit,
independently on the way the chiral symmetry is realized:
\begin{equation}
{(m_u+m_d)(1~\rm GeV)} \leq 15.7~ {\rm MeV}. 
\end{equation}
Therefore, one can conclude the range of values for the sum of light quark masses:
\beq
9.4~{\rm MeV}\leq {(m_u+m_d)(1~\rm GeV)} \leq 15.7~ {\rm MeV}, 
\end{equation}
which is independent on the nature of the realization of chiral symmetry
breaking.\\
Combining the lower bound, obtained within a GMOR realization
of the chiral symmetry breaking, with the ratio of the light quark
masses from chiral perturbation theory, namely \cite{LEUT}:
\beq
\frac{m_s}{1/2(m_u+m_d)}=25.7\pm 2.6,
\eeq
one obtains:
\beq
m_s(1~\rm GeV)\geq (121\pm 12)~\rm{MeV}.
\eeq
The lower bounds in Eqs. (10) and (14) are in agreement with the direct
determinations of the quark masses from the pseudoscalar sum rule
\cite{PSEUDO,EDUAR},
but, disagree with recent extrapolations from Monte Carlo
simulations \cite{LATT}. These lower bounds also agree with the 
lower bounds from the pseudoscalar \cite{PSEU,EDUAR} and scalar 
\cite{SCALAR} sum rules. However, we expect that, in view
of the value of the chiral condensate obtained from the combined nucleon and
$B^*$-$B$ sum rules, the lower bounds in Eqs. (10) and (14), though
interesting, is relatively weak compared to the real estimate. Indeed,
the value of the quark condensate in Eq. (6) from these combined sum rules
leads to the estimate:
\begin{equation}
{(m_u+m_d)} (1~{\rm GeV})\simeq 15~{\rm MeV}~,~~~~~~~~~~~~~~
m_s(1~\rm{GeV})\simeq 182~\rm{MeV}~,  
\end{equation}
which is in good agreement with the pseudoscalar
\cite{PSEU,PSEUDO,EDUAR}, and earlier scalar \cite{SCALAR,CHET} sum rule results.
The above range of values of the strange quark mass also agrees with 
the recent estimate of $m_s$(1 GeV)=$ (197\pm 29)$ MeV
from the $e^+e^-\rar$ hadrons \cite{MS} and the preliminary
result of about 200 MeV from $\tau$-decay data \cite{ALEPH}. It should be noted
that some of these previous analyses, especially \cite{EDUAR} starts
from more general principles than our investigation and therefore the limits
are somewhat weaker.
\section*{Acknowledgements}
Some discussions on this work have been initiated during the QCD 97
Euroconference held in Montpellier. H.G.Dosch would like to thank the CNRS
for financial support and the LPMT of Montpellier for the hospitality offered
to him.
\vfill\eject

\end{document}